\documentclass[10.5pt,a4paper]{article}
\usepackage{CJK}

\usepackage[none]{hyphenat}
\usepackage{mathrsfs}
\usepackage{amsmath}
\usepackage{multicol}
\usepackage{graphicx}
\usepackage{CJK}
\usepackage{multirow}
\usepackage{fancyhdr}
\usepackage{graphics}
\usepackage{tabularx}
\usepackage[parfill]{parskip}
\usepackage[doi=false,url=false,sorting=none]{biblatex}
\begin{document}


\title{Fast dynamic wavefront correction for multi-photon microscopy with a high resolution MEMS phase-only modulator}

\maketitle

\noindent {\bf\normalsize{Juan David Muñoz-Bolaños\textsuperscript{\textdagger}, Eva Ernst\textsuperscript{\textdagger}, Maria Borozdova, Florian Harrasser, Daniel Grünbacher, Alexander Knapp, Simon Moser$^{*}$, Monika Ritsch-Marte and Alexander Jesacher}}


\noindent{\textsuperscript{\textdagger}\footnotesize{These authors contributed equally to this work.}}\\
\noindent{\footnotesize{Medical University of Innsbruck, Institute of Biomedical Physics, Innsbruck, 6020, Austria}}\\


\noindent{\small{\textbf{Abstract.} Multi-photon microscopy is a powerful technique for deep-tissue imaging, providing high spatial resolution at increased penetration depth. Nevertheless, imaging remains largely restricted to superficial tissue layers well below 1 mm. Adaptive optics based on indirect wavefront sensing can significantly extend the accessible imaging depth, but their iterative operation is typically time-consuming and therefore limits the correction of strong, spatially complex aberrations. Until recently, this limitation was partly due to the lack of high-speed phase modulators offering sufficiently large numbers of actuators.
Recent technological advances have addressed this bottleneck with the emergence of megapixel phase modulators operating at kilohertz rates. Here, we demonstrate wavefront correction in multi-photon microscopy using a high-speed phase-only MEMS modulator combined with a rapidly converging scatter-compensation algorithm. We experimentally correct complex aberrations comprising 144 spatial modes in less than one second, resulting in signal enhancements exceeding a factor of two.
Benchmarking against a fast liquid-crystal spatial light modulator reveals a sixfold increase in correction speed. We further demonstrate adaptive optics imaging in mouse brain tissue using both two-photon and three-photon excitation fluorescence microscopy. These results indicate that high-resolution MEMS-based spatial light modulators enable efficient indirect wavefront sensing and represent a promising platform for high-speed wavefront shaping in non-linear microscopy.}}\vspace*{4mm}

\vspace*{4mm}

\noindent{\small{$^\ast$Simon Moser, E-mail:
simon.moser@i-med.ac.at}}\vspace*{10mm}

\sloppy{}


\section{Introduction}

\noindent Multi-photon fluorescence microscopy (MPM)~\cite{Denk1990, Helmchen2005} has emerged as a powerful microscopic method to image deep inside thick biological tissues. Its high penetration depth arises from the use of near-infrared (NIR) excitation wavelengths, combined with the method’s intrinsic optical sectioning capability. 
However, even though MPM techniques can penetrate deeper than linear fluorescence methods such as confocal or light-sheet microscopy, optical aberrations and scattering still limit the imaging depth to significantly below one millimetre in most cases~\cite{xu2024multiphoton}.
One path to increase the penetration depth in microscopy is the use of adaptive optics (AO)~\cite{marsh2003practical, debarre2009image, ji2010adaptive}. 
AO uses spatial light modulators (SLMs) to shape the excitation laser such that scattering induced by the tissue is compensated and a compact focal spot forms inside the sample. Methods to identify aberrations imposed by the tissue can be separated into \emph{direct} methods, which measure the wavefront using a guide star, and \emph{indirect} methods that systematically perform sequences of measurements to identify the aberration . 
Direct methods usually employ Hartmann-Shack sensors which capture wavefronts in single measurements~\cite{aviles2011measurement}. Because of the necessity to image the signal from the focus back through the scattering medium, direct methods are increasingly difficult to use in scattering environments. Nevertheless, recent research showed remarkable progress using infrared dyes and optimized measurement configurations~\cite{liu2019direct, chen2021high}.
Indirect methods are more robust to scattering but require a set of test measurements and are therefore typically slower. 
Finally, \textit{interferometric focus sensing techniques} are a specific class of indirect methods that have been developed for strongly scattering samples~\cite{Tang2012, Papadopoulos2016, May2021, qin2022deep, blochet2023fast}. 
These techniques split the excitation beam into a modulated part that is phase-stepped with respect to a constant reference part. Zonal wavefront sensing~\cite{ji2010adaptive} can also be viewed as a member of this family. 

In 2020, some of us have developed \textit{dynamic adaptive scattering compensation holography} (DASH)~\cite{May2021, Nam2025}, an interferometric technique related to F-SHARP~\cite{Papadopoulos2016} that has a simpler hardware implementation and completes the wavefront measurement in fewer iterations. 
DASH uses only a single pixelated SLM and does not require any other elements such as beam splitters, phase-steppers or scanning modules. 
Furthermore, DASH is practically 'alignment-free'. This means that the often cumbersome process of aligning the SLM mask with the exit pupil of the objective lens is unnecessary, as the correction pattern forms automatically at the optimal position during the measurement process.
This simplicity, however, comes at the cost of the SLM having to take on the tasks of these additional elements, requiring to update the displayed diffractive patterns several hundred times during one single DASH run.
Therefore, unlocking the full potential of DASH requires a fast diffractive SLM, ideally with switching times on the order of microseconds. 
For AO implementations in MPM, liquid crystal on silicon (LCoS) SLMs have long been the most attractive choice due to their high resolution and capability to provide continuous phase-only modulation. The typical response time of LCoS SLMs, however, lies within tens of milliseconds, which can be improved down to a few milliseconds~\cite{Thalhammer2013}, which is still two orders of magnitude away from enabling optimal performance of DASH. Recently, Texas Instruments has been developing phase light modulators (PLMs) based on micro electro mechanical systems (MEMS) technology, offering switching times of 50 \textmu s at comparable resolutions to LCoS SLMs~\cite{Bartlett2019,Bartlett2021,Douglass2022}. 

In this work we present an implementation of DASH using fast MEMS phase only modulators (PLMs). We introduce modifications of DASH that account for the specific control scheme of the PLM and provide instructions for its implementation. We further compare the performance of the PLM to a fast LCoS device in terms of speed and quality using fluorescently labelled beads and demonstrate the capability of NIR-PLMs by scatter compensation for two-  and three-photon excited fluorescence imaging of microglia cells in mouse brain tissue. We experimentally demonstrate that our implementation enables the correction of 144 spatial modes in less than a second, which outperforms the capabilities of conventional liquid crystal based setups by a factor of six. 

\section{Methods}

\subsection{Phase Light Modulator}

The devices used for wavefront shaping are .67" PLMs from Texas Instruments. The first device (VIS-PLM) has $1358 \times 800$ individually addressable mirrors with a pixel pitch of 10.8 \textmu m designed for the visible wavelength range (405-650 nm) allowing 16 discrete phase steps (4-bit). The NIR (700-1550 nm) device (FFE Alpha) with $904 \times 800$ addressable mirrors (pixel pitch of 10.8 \textmu m $\times$ 16.2 \textmu m) having 32 discrete phase steps (5-bit) is the second device (NIR-PLM) that we implement for scattering compensation. With a switching time of only 50 \textmu s the PLMs offer comparable resolutions to a LCoS SLM at much higher switching speed. Furthermore, as the phase shift is imprinted by displacing the micro-mirrors it precludes problems such as polarization sensitivity and conversion as well as pixel cross-talk present in LCoS SLMs~\cite{Moser2019}.

In order to drive the current PLM the hardware and software developed by Texas Instruments for DMDs has been adapted to allow addressing the individual pixels of the PLM. For a concise description and characterization of the PLM we refer to the work of Rocha et al.~\cite{Rocha2024}. 

We implemented the data streaming using the DisplayPort interface of the PLM, which allows to transfer batches of 24 holograms at 60 Hz with a hologram display rate of 1440 Hz. To transfer the data from the host-PC to the EVM we developed custom software based on OpenGL and CUDA controlled via Python. The code is made available at~\cite{dashplmgit}. 

\subsection{Batched DASH}

The original DASH algorithm sequentially applies a set of $K$ plane wave modes $M_k$, and phase-steps them using $L$ steps $\theta_l$ against a continuously improving correction pattern $C_k$ serving as reference. Both fields are simultaneously shaped by a single phase hologram. The complex-valued holograms thus have the following form
\begin{align}
    h_{k, l} = \sqrt{1-f} \frac{C_k}{|C_k|} + \sqrt{f} M_k \exp(j \theta_l).
\end{align}
where $f$ denotes the power fraction contained in the test modes (in our case is chosen in between $0.1$ and $0.3$). For the complex DASH version $h_{k, l}$ is formed~\cite{Sohmen2024, Nam2025}  by means of complex modulation while for the phase-only version $\arg(h_{k,l})$ is displayed. For every hologram displayed an n-photon-excited fluorescence signal $I^\mathrm{n-ph}_{k,l}$ is recorded, from which the complex weight $w_k$ is identified. The optimal phase value for plane wave $M_k$ is calculated using $\phi_k = \arg(a_k)$ with
\begin{align}
    a_k = \frac{1}{L}\sum_l \sqrt[n]{I^\mathrm{n-ph}_{k,l} }\exp(j \theta_l)
\end{align}
while the amplitude is calculated using~\cite{Nam2025}
\begin{align}
    |w_k| = \sqrt{\frac{1}{2f}\big(\bar{S}_k - \sqrt{\bar{S}_k^2 - 4 |a_k|^2}\big)}
\end{align}
with $\bar{S}_k$ denoting the average of $\sqrt[n]{I^\mathrm{n-ph}_{k,l}}$ over $L$ phase steps. The correction pattern is then updated according to
\begin{align}
    C_{k} = C_{k-1} + |w_k|\exp(j \phi_k).
\end{align}

The display of the PLM is batched to 24 holograms, which upon transfer are displayed sequentially. 
We execute DASH in a 'batched mode', i.e. we update $C$ only after measuring a number of modes fitting into a set of 24 holograms. While such a batched process is expected to have a negative impact on the convergence speed of DASH, our simulations and experiments show that the effect is practically negligible if a single batch comprises only up to 8 modes (see Appendix). Of note, the performance of DASH converges to that of the related technique IMPACT~\cite{Tang2012} if a single batch contained all $N$ modes. 

For pattern calculation and update as well as the bit-packing routines necessary for the PLM we leverage GPU acceleration using CUDA and OpenGL for rendering. For the wavefront correction using DASH with the PLM operating at 1440 Hz we continuously display and update the patterns every 16.6 ms. We use a Nvidia RTX 3060 GPU to calculate and process the holograms for the PLM. When operating the PLM at a resolution of 1280x800, the time for updating the correction pattern, calculation of 24 subsequent holograms as well as bit-packing takes around 2 ms, which is well below the 16.6 ms time interval set by the DisplayPort interface. We utilize buffer-sharing between CUDA and OpenGL to minimize copy operations. However, in the pipeline for data transfer from the host PC to the PLM device over the DisplayPort we incur a latency stemming from the internal display queue of the GPU, which needs to be taken into account for correctly assigning each measured signal to its hologram. To identify the delay present in the pipeline from host PC to PLM we display a sequence of gratings with increased modulation depth and record the PMT signal on test samples such as fluorescently labelled beads. Simulations investigating the combined effect of batching and delay on the speed of convergence of DASH are provided in the Appendix.

\subsection{Experimental Setup}

\begin{figure}[h!]
    \centering
    \includegraphics[width=\linewidth]{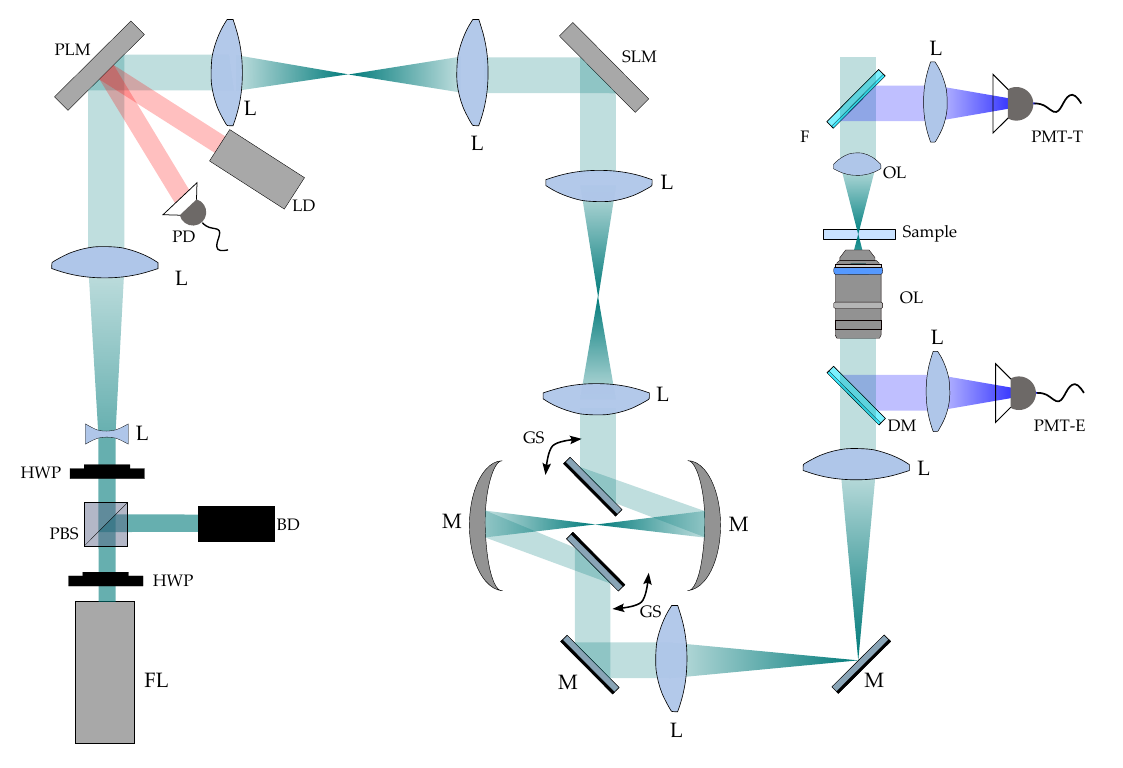}
    \caption{\textbf{Multi-photon imaging setup:} Light from femtosecond lasers (MaiTai Deep See and White Dwarf) is guided into a power control consisting of a half wave plate, polarizing beam splitter and beam dump. The beam is expanded through a telescope, guided onto the PLM and relayed to the SLM, the galvo scanners and the back-focal plane of the objective lens. The generated multi-photon fluorescence signal is then collected and guided onto photomultiplier tubes in epi- and trans-direction. The light signal generated from a laser diode guided onto the PLM is recorded from a photodiode is used for synchronization. FL: Femtosecond laser (MaiTai DeepSee/White Dwarf), HWP: half-wave plate, PBD: polarizing beam splitter, BD: beam dump, PLM: MEMS phase light modulator, SLM: LCoS spatial light modulator, L: lens, PD: photo diode, LD: laser diode, M: mirror, GS: galvo scanner, DM: dichroic mirror, OL: objective lens, CL: condenser lens, F: notch-filter, PMT-T: photomultiplier tube for trans detection, PMT-E: photomultiplier tube for epi detection. }
    \label{fig:exp_setup}
\end{figure}

Figure~\ref{fig:exp_setup} depicts the experimental setup for scattering compensation in multi-photon microscopy. For two-photon imaging, we use a wavelength tunable Titanium-Sapphire laser ('MaiTai DeepSee' from Spectra-Physics\textsuperscript{\textregistered}), while we use a 'White Dwarf' laser from Class5 for three-photon imaging at an excitation wavelength of 1300 nm. The beams of both lasers are separately expanded, combined via a dichroic beam splitter and finally directed onto the PLM panel. 
The PLM is further imaged onto the LCoS SLM using a 4f relay and both planes are conjugated to a pair of galvanometric scanning mirrors and the exit pupil of the microscope objective (25X, 0.95 NA, Leica HC FLUOTAR water immersion). The epi-collected fluorescence signal is reflected by a dichroic mirror and directed into a photomultiplier tube (H10769PA-40 from Hamamatsu Photonics). The trans-collected signal is collected by a condenser lens and guided to an additional photomultiplier tube.

The PLM is initialized using the DLP Lightcrafter DLPC900 GUI from TI and operated in the Video Pattern Mode for continuous display. In this mode the PLM can be configured to send a TTL signal after every pattern update every 694 \textmu s, which is used for synchronization with a PMT for signal acquisition. The signal acquisition is controlled by a National Instruments PXIe-7846R FPGA system programmed in LabView. 

Since the PLM is continuously running, it is necessary to synchronize the starting point of the signal acquisition for the DASH measurement series. We perform this by using a separate diode laser and an amplified fast photodiode. Immediately before each DASH run we display a grating, which diffracts the diode laser beam onto the photodiode and synchronize the data acquisition to a TTL pulse derived from the photodiode signal.  

\section{Results}

\subsection{Performance comparison of PLM and LCoS SLM}

The PLM used for the experiment described in this section is designed for operation with optical wavelengths in the range from 405-650 nm. Because it has insufficient stroke for the NIR range between 800-900 nm we implement DASH in an off-axis configuration by adding a vertical blazed phase grating with a period of 10 pixels to all the patterns displayed on the VIS-PLM. The first diffraction order can be given any phase shift regardless of the PLMs stroke, however, this comes at the cost of a reduced light efficiency and wavelength dispersion. For our particular implementation, the VIS-PLM had a first-order diffraction efficiency of $\sim20$\% at 840 nm for a blazed grating with a pixel period of 10. 

\begin{figure}[h!]
 \centering
        \includegraphics[width=\textwidth]{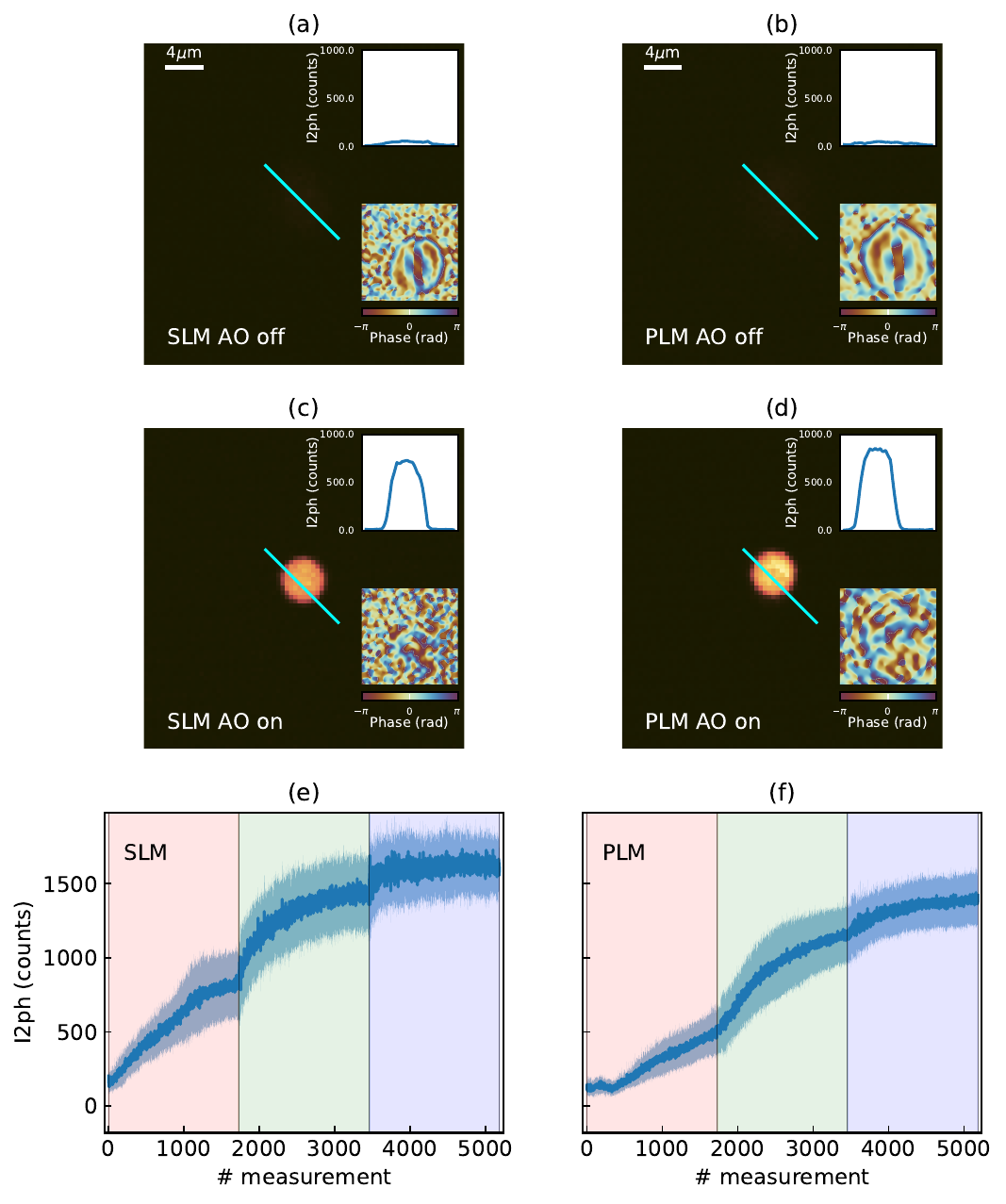}
 \caption{\textbf{DASH with LCoS SLM and PLM: } Panels (a) and (b) show two-photon excited fluorescence images of fluorescently labelled beads placed under scattering tape with only the correction mask applied for system aberrations for the SLM and PLM, respectively. In (c) and (d) the images after the DASH correction run is shown for both devices, while (e) and (f) depict the average signal enhancement for 8 individual DASH correction runs. }
\label{fig:comparison_enhancement}
\end{figure}

To benchmark the implementation of the batched version of DASH using the PLM we compare it to an LCoS SLM of comparable resolution (Meadowlark HSP1920-500-1200-HSP8 with 1920x1152 pixels). As test samples we used 4~\textmu m fluorescent beads (TetraSpeck\texttrademark ~microspheres from ThermoFisher Scientific) hidden under a 170~\textmu m thick coverslip with a layer of matt scattering tape (Scotch\textregistered ~Magic\texttrademark ~tape). First, we compare the performance of both devices in terms of enhancement, which is shown in  Fig.~\ref{fig:comparison_enhancement}. The enhancement curves in Fig.~\ref{fig:comparison_enhancement} for the SLM and PLM are shown in (e) and (f), respectively. In both DASH runs $N_i=3$ iterations were performed with $L=3$ phase steps and a number of $K=24^2=576$ plane-wave modes were corrected. The solid curves show the average over four runs performed for different beads. The lighter coloured ribbons visualize the respective standard deviation. The DASH optimization for the PLM is performed batch-wise and with updates delayed by two frames (48 holograms), while for the LCoS SLM the correction pattern is updated immediately after the $L=3$ measurements. To establish comparable signal levels for both experiments, we adapted the respective signal integration times, which we set to 150~\textmu s for the PLM experiment and 200~\textmu s for the LCoS SLM experiment.

\begin{figure}[h!]
    \centering
    \includegraphics[width=0.5\linewidth]{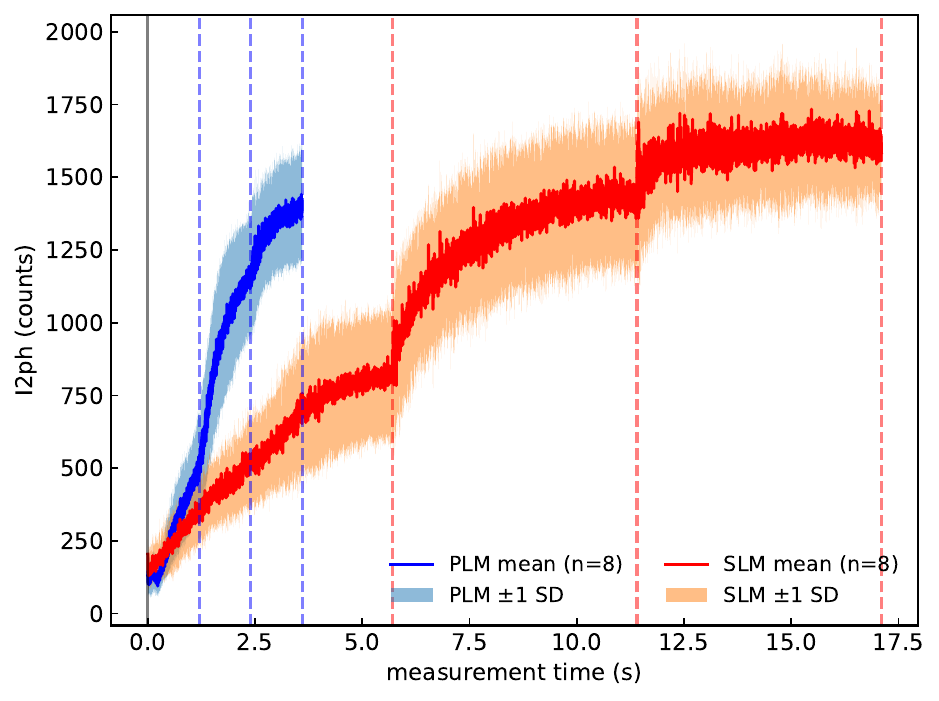}
    \caption{\textbf{Speed comparison of LCoS SLM and PLM:} In blue the DASH enhancement curve of the PLM is shown, while red shows the enhancement curve for the LCoS SLM. Each vertical dashed line marks the end of one optimization iteration. For both devices an equal number of measurements is performed.}
    \label{fig:comparison_speed}
\end{figure}

In Fig.~\ref{fig:comparison_speed} we see DASH enhancement curves plotted against the measurement time. The PLM essentially performs the same measurements at six times the speed of the LCoS SLM.

\subsection{Biological imaging}

For the results presented in this section, the NIR PLM device was used for wavefront shaping in the imaging of GFP labelled microglia cells in a fixed traverse mouse brain tissue slice of about 1 mm thickness. For two- and three-photon imaging excitation, wavelengths of $920$ nm and $1300$ nm were used, respectively. The NIR PLM used in this work is a prototype (alpha) early in development, which exhibits imperfections such as dead pixels and state-dependent tilts, which tend to increase at higher piston states. As the device is designed for operation up to 1630 nm we utilized the bottom 20 and 28 states for two- and three-photon excitation, respectively.  

\begin{figure}[h!]
 \centering
    \includegraphics[width=\textwidth]{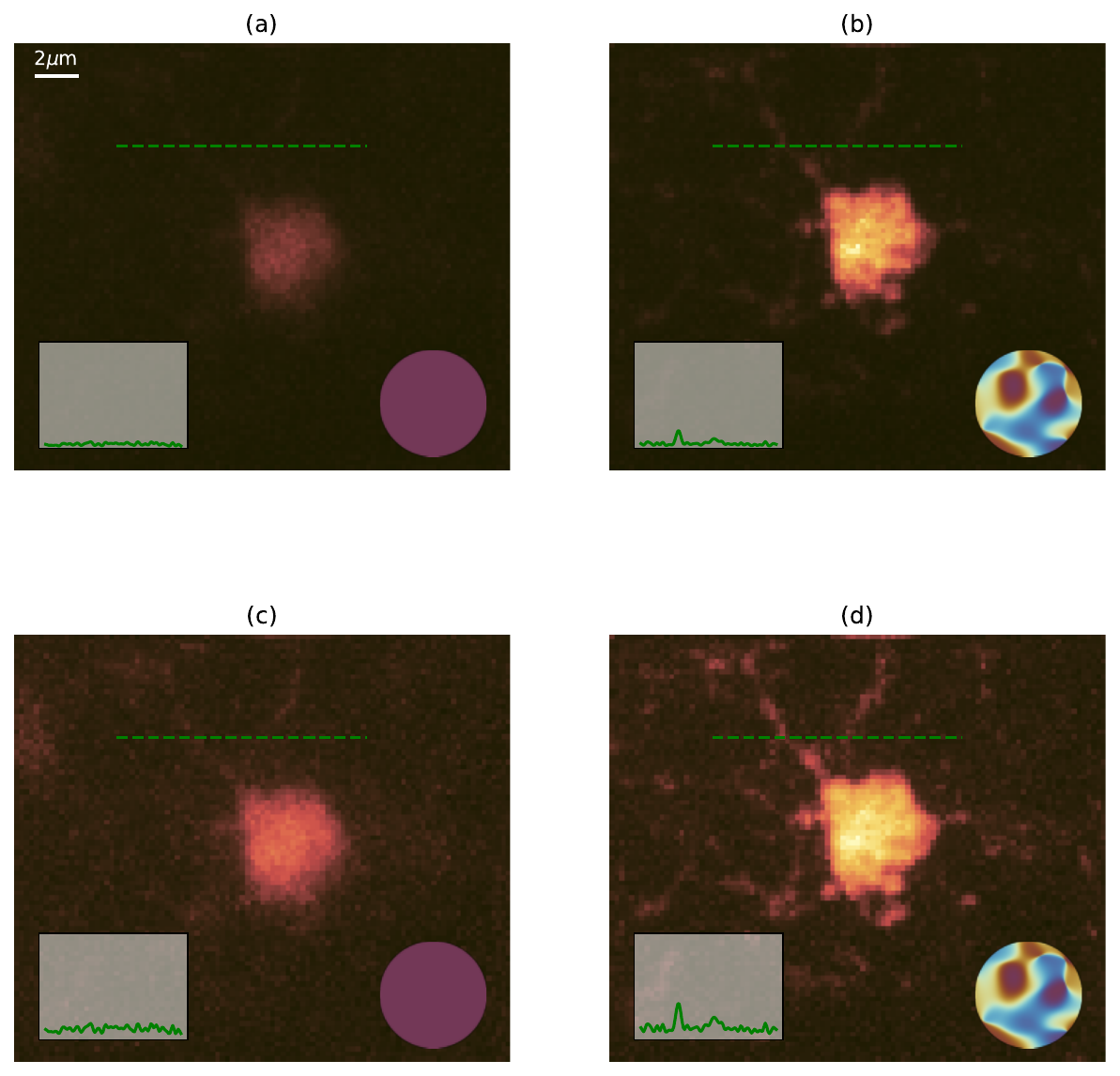}
 \caption{\textbf{Scattering compensation in two-photon imaging:} In (a) and (b) the maximum intensity projections of uncorrected and corrected two-photon image z-stacks recorded at a depth of 150 \textmu m are shown, respectively. The panels (c) and (d) show the same images at a logarithmic scale.}
\label{fig:bio_2ph}
\end{figure}

Figure~\ref{fig:bio_2ph} shows two-photon images taken at a depth of about~150 \textmu m without correction (a,c) and after a DASH correction run (b,d). The top panels (a,b) depict maximum projections of small z-stacks taken around the cell, whereas the bottom panels (c,d) show the same images in logarithmic scale. We corrected $K=144$ plane-wave modes using $L=3$ phase steps in three iterations, which results in a total of 1296 measurements with a total correction time of $0.9$ s. The glial cell is enhanced by a factor of $\sim 2.5$ in the corrected images, clearly revealing finely structured processes which are not visible in the uncorrected image stacks.   

\begin{figure}
    \centering
    \includegraphics[width=\linewidth]{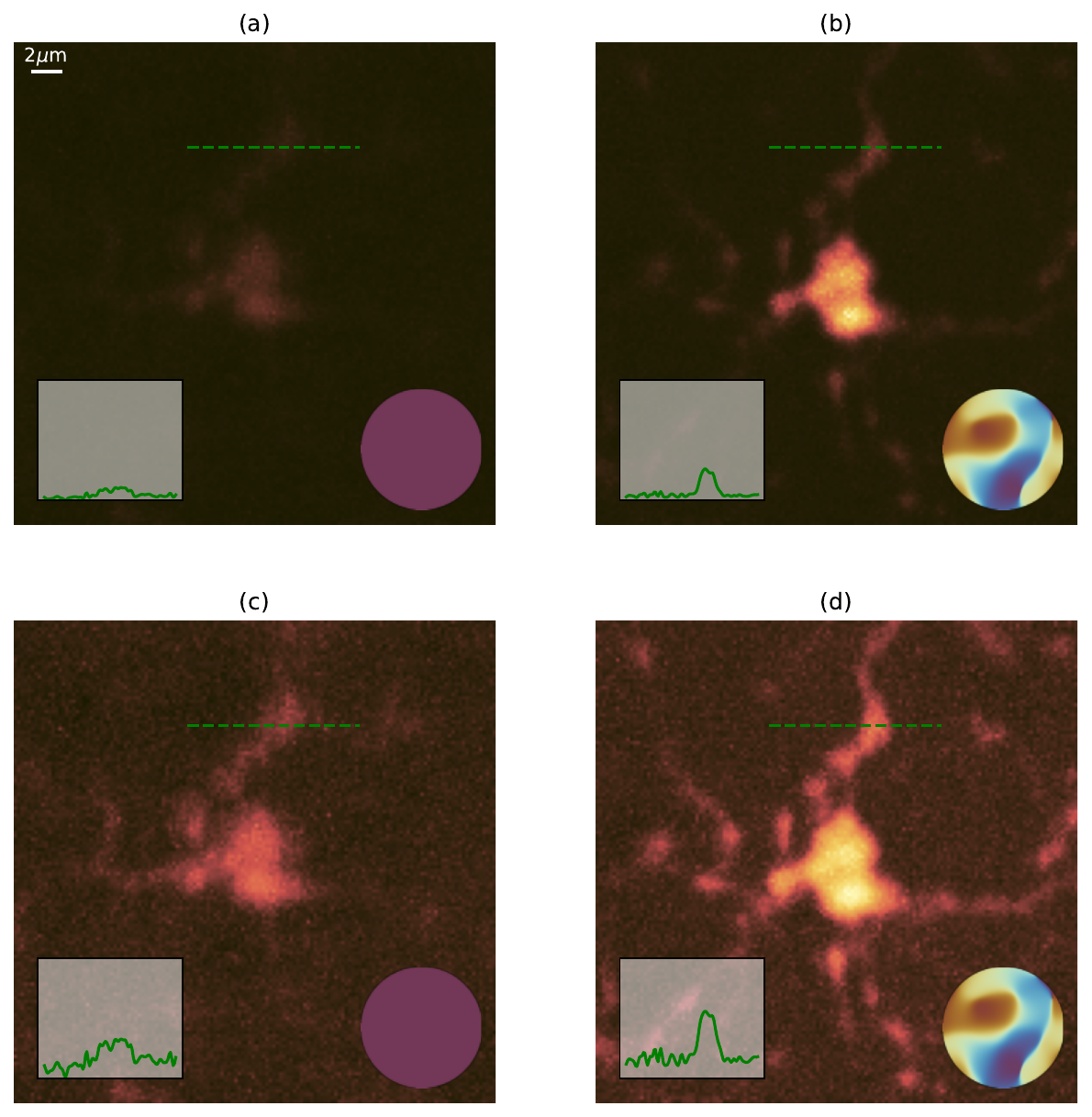}
    \caption{\textbf{Scattering compensation in three-photon imaging:} Panels (a) and (b) show a single microglia cell at 450 \textmu m depth inside fixed mouse brain tissue, where (a) shows the uncorrected and (b) the corrected three-photon image. Panels (c) and (d) show (a) and (b) in logarithmic scaling, respectively.}
    \label{fig:bio_3ph_1}
\end{figure}

Figure~\ref{fig:bio_3ph_1} shows results from scattering compensation in three-photon fluorescence imaging at a depth of $\sim 450$ \textmu m. In (a) and (b) the maximum intensity projections of the uncorrected and corrected images are shown, respectively. Panels (c) and (d) show the same images as (a) and (b) in logarithmic scaling. Compared to the uncorrected images (a,c), the enhanced images (b,d) allow one to identify cellular processes that are otherwise poorly visible or not visible at all. 

\begin{figure}
    \centering
    \includegraphics[width=\linewidth]{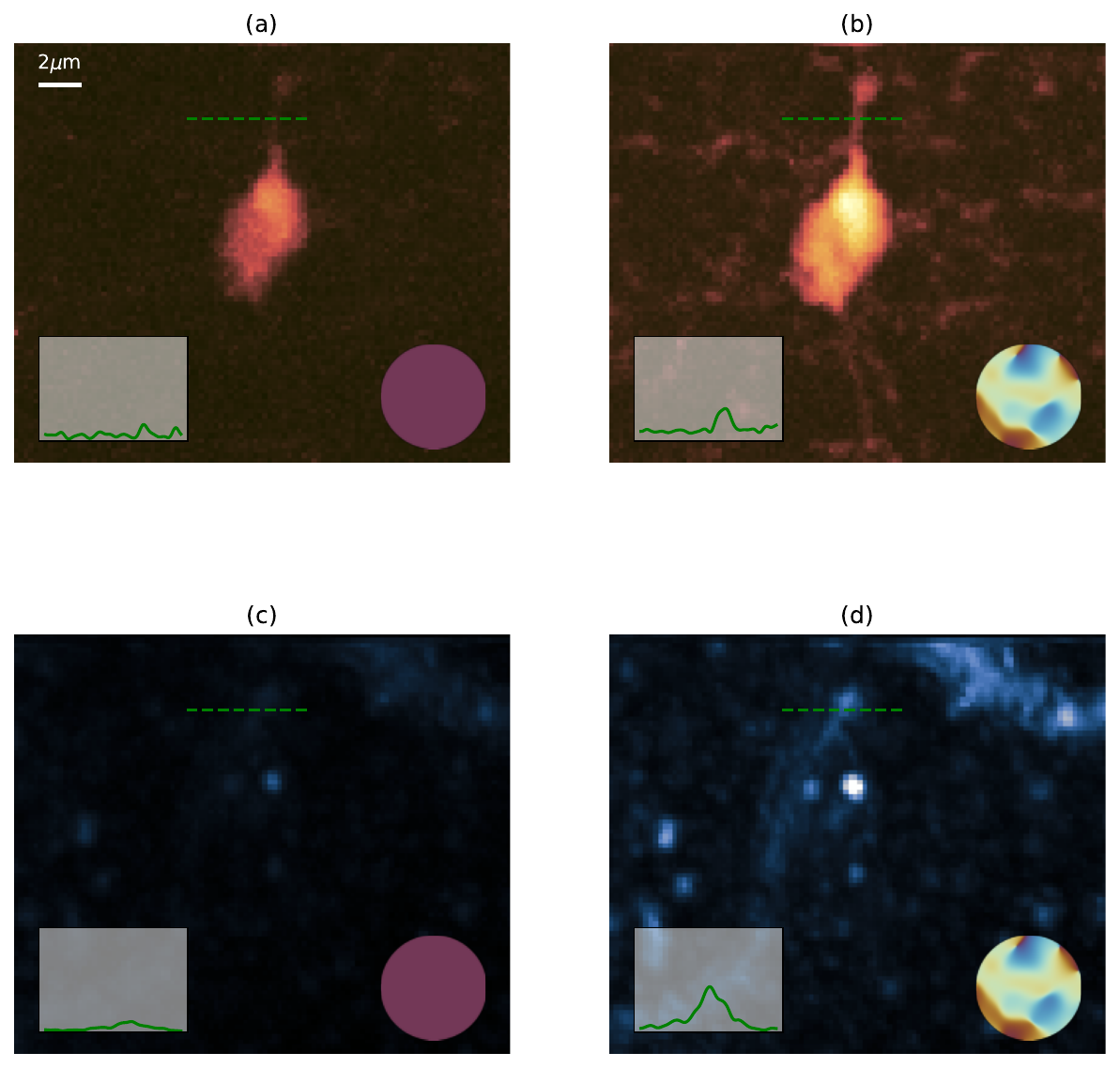}
    \caption{\textbf{Scattering compensation in combined three-photon fluorescence and third-harmonic generation imaging:} Panels (a) and (b) show logarithmically scaled maximum intensity projections of stacks over a scan region of a microglia cell at 450 \textmu m depth in hippocampal mouse brain tissue, where (a) shows the uncorrected and (b) the corrected three-photon image. Panels (c) and (d) show maximum intensity projections of third harmonic generation image stacks collected with the correction obtained from the fluorescent signal.}
    \label{fig:bio_3ph_2}
\end{figure}

Figure~\ref{fig:bio_3ph_2} shows results from scattering compensation in combined three-photon fluorescence and third harmonic generation imaging at a depth of $\sim 460$ \textmu m.  The corrected THG signal was obtained with the scattering compensation mask computed from the three-photon fluorescence signal. In (a) and (b) the maximum intensity projections of the uncorrected and corrected images are shown, respectively. Panels (c) and (d) show the third harmonic generation signal over the same region as the images as (a) and (b).Compared to the uncorrected images (a,c), the enhanced images (b,d) also show a significant signal enhancement for THG.

\clearpage
\section{Discussion and Conclusion}

In this work we performed wavefront shaping for aberration compensation in MPM by performing DASH with high-resolution MEMS phase only light modulators. The PLMs provide hologram refresh rates of 1440 Hz at megapixel resolution. Given that multi-photon microscopy typically operates with integration times of up to 100 \textmu s,  the PLM with its switching time of 50 \textmu s is the first device that grants access to this time-regime at high resolution. Compared to LCoS SLMs, which are limited to hologram switching times in the millisecond range, the PLM reduces the dead time-frame during which the sample is exposed without data acquisition by a factor of 20 or more.

A limitation of the VIS-PLM is its 4-bit phase resolution, providing only 16 addressable phase levels. Compounded with deviations from linearity in the level-to-phase relationship, the VIS-PLM might have limited applicability for high-precision beam shaping. While the NIR device offers 5-bit depth and higher linearity, it does so at the price of a smaller spatial resolution. However, for the purpose of indirect wavefront-sensing in MPM we could identify no significant degradation in performance resulting from the limited number of available phase levels. 

We provided an adaptation for the operation of DASH with signal batches and compared the performance of the PLM to a state of the art LCoS SLM for fluorescently labelled beads in comparable imaging conditions and found the PLM to provide a similar signal enhancement at 6 times the speed. We further performed aberration correction on microglia cells in mouse brain tissue with a NIR-PLM with two- and three-photon excitation, which showed similar signal enhancements to other devices. With further developments and refinements of the NIR MEMS technology, we believe this platform will become highly attractive for high-speed wavefront shaping and adaptive optics in MPM. 

\vspace*{4mm} \noindent \emph{Disclosures}\vspace*{4mm}

\noindent The authors declare no conflict of interests.

\vspace*{4mm} \noindent \emph{Acknowledgments} \vspace*{4mm}

\noindent We would like to thank David B. Phillips and Jose Carlos do Amaral Rocha for help in the instrumentation of the PLM. We would like to thank Kai Kummer for providing the mouse brain slices and Werner Kirchler for his support in constructing the multi-photon imaging platform. This research was supported by the following research Grants: Austrian Science Fund (FWF) DOC. 110-B: Image-guided Diagnosis and Therapy; Austrian Science Fund (FWF) (10.55776/P36687).

\printbibliography

\section*{Appendix}

\begin{figure}[h!]
    \centering
    \includegraphics[width=\linewidth]{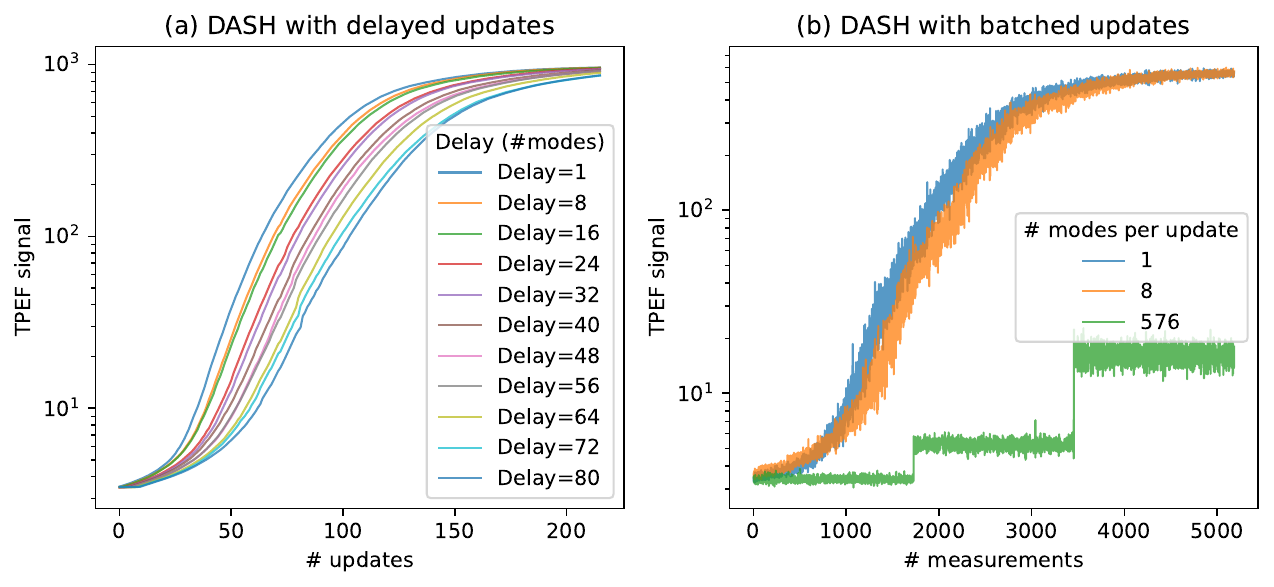}
    \caption{\textbf{DASH with delay and batching:} In (a) the DASH enhancement curves are shown, where updates are computed and applied every 8 modes (24 holograms) with a variable delay. The figure in (b) depicts DASH enhancement curves where updates to the correction patterns are performed with a variable batch size.}
    \label{fig:batch_delay_sim}
\end{figure}

Figure~\ref{fig:batch_delay_sim} (a) shows DASH runs where the updates of the correction patterns incur a delay of a number of modes. Here a number of phase steps $N_p=3$ is assumed, which yields a delay of multiples of 8 modes, which can be probed during a single set of frames sent to the PLM. In Fig.~\ref{fig:batch_delay_sim} (b) the influence of batching on the signal enhancement progression is displayed, where the correction pattern is updated after probing a number of modes. The curve where all $576$ are probed before updating the correction pattern is equivalent to IMPACT/F$-$SHARP. Both batching and delay do underestimate the mode amplitudes compared to DASH with a delay of $0$ and with updating the pattern after each mode is probed. This underestimation results in enhancement curves that lag behind.

\end{document}